\documentclass[showpacs,aps,prd,preprint,superscriptaddress]{revtex4-1}
\usepackage[colorlinks=true, pdfstartview=FitV, linkcolor=red,citecolor=blue, urlcolor=blue,
bookmarks=true, bookmarksnumbered=true, breaklinks]{hyperref}
\usepackage[dvipdfmx]{graphicx}

\usepackage{amsmath,amssymb,bm,ascmac,color,longtable,mathrsfs}

\usepackage{slashed}

\allowdisplaybreaks[1]

\usepackage{color}

\usepackage[normalem]{ulem}

\def\simge{\mathrel{
   \rlap{\raise 0.511ex \hbox{$>$}}{\lower 0.511ex \hbox{$\sim$}}}}
\def\simle{\mathrel{
   \rlap{\raise 0.511ex \hbox{$<$}}{\lower 0.511ex \hbox{$\sim$}}}}
\def\bigs{\mathrel{
   \rlap{\raise 0.531ex \hbox{$>$}}{\lower 0.531ex \hbox{$<$}}}}

\allowdisplaybreaks[3]
\allowdisplaybreaks

\usepackage{braket}
\def\vector#1{\mbox{\boldmath $#1$}}

\newcommand{\iiid}{V_{\mathrm{III2}}}
\newcommand{\iiit}{V_{\mathrm{III3}}}
\newcommand{\conf}{V_{\mathrm{conf}}}

\newcommand{\iti}{{\bf 1}}
\newcommand{\san}{{\bf 3}}
\newcommand{\sanbar}{{\bf \bar{3}}}
\newcommand{\roku}{{\bf 6}}

\newcommand{\hati}{{\bf 8}}

\newcommand{\ccbar}{c\bar{c}}
\newcommand{\uds}{uds}
\newcommand{\cc}{c\bar{c}}
\newcommand{\po}{P_{cs\hati}}
\newcommand{\pop}{P'_{cs\hati}}
\newcommand{\poa}{P^{\ast}_{cs\hati}}
\newcommand{\ps}{P_{cs\iti}}
\newcommand{\psp}{P'_{cs\iti}}
\newcommand{\psa}{P^{\ast}_{cs\iti}}

\begin{document}

\title{Flavor-singlet charm pentaquark}

\author{Yoya~Irie}
\affiliation{Department of Physics, Tokyo Institute of Technology, Tokyo 152-8551, Japan}
\author{Makoto~Oka}
\email[]{oka@th.phys.titech.ac.jp}
\affiliation{Department of Physics, Tokyo Institute of Technology, Tokyo 152-8551, Japan}
\author{Shigehiro~Yasui}
\email[]{yasuis@th.phys.titech.ac.jp}
\affiliation{Department of Physics, Tokyo Institute of Technology, Tokyo 152-8551, Japan}

\begin{abstract}
A new type of charm pentaquark $P_{cs}$ with quark content $c\bar{c}uds$ in light-flavor singlet state is studied in 
the quark model.
This state is analogous to the $P_{c}$ with $c\bar{c}uud$ in light-flavor octet, which was observed in LHC in 2015.
Considering various combinations of color, spin and light flavor as internal quantum numbers in $P_{cs}$,
we investigate the mass ordering of the $P_{cs}$'s by adopting both the one-gluon exchange interaction and the instanton-induced interaction in the quark model.
The most stable configuration of $P_{cs}$ is identified to be total spin $1/2$ in which the $c\bar{c}$ is combined to be color octet and spin $1$, while the $uds$ cluster is in a color octet state.
The other color octet configurations, the total spin $1/2$ state with the $c\bar{c}$ spin 0 and the state with total spin $3/2$ and $c\bar{c}$ spin 1, are found as excited states.
We also discuss possible decay modes of these charm pentaquarks.
\end{abstract}


\maketitle

\section{Introduction}

Studying exotic hadrons, so called $X$, $Y$, $Z$, is one of the most interesting topics in the present hadron physics~\cite{Swanson:2006st,Voloshin:2007dx,Brambilla:2010cs,Brambilla:2014jmp,Chen:2016qju,Hosaka:2016pey,Lebed:2016hpi,Esposito:2016noz}.
In 2015, a new type of exotic hadron, a pentaquark with hidden charm $P_{c}$, was observed in LHC experiment~\cite{Aaij:2015tga}.
$P_{c}$ is considered to be $c\bar{c}uud$ as a minimal quark configuration, and hence this is the first discovery of pentaquark including charm quarks.
$P_{c}$ was observed in $J/\psi p$ channel in the weak decay from $\Lambda_{b}$ baryon, and the two states were identified: $P_{c}(4380)$ with mass $4380\pm30$ MeV and decay width 205 MeV, and $P_{c}(4450)$ with mass $4449.8 \pm 3.0$ MeV and decay width $39 \pm 20$ MeV.
To identify the internal structure of $P_{c}$ is the most fundamental problem currently.
Although there are a large number of theoretical studies about charm pentaquarks in literature, however, there is not yet conclusive picture about the structure of $P_{c}$.

Let us briefly summarize studies of charm pentaquarks.
As an early work, existence of charm pentaquark was pointed out in the framework of the Skyrmion model, where $\eta_{c}$ meson is bound to the hedgehog configuration of pion~\cite{Gobbi:1992cf}.
Afterwards, hadron molecule model was analyzed in Refs.~\cite{Hofmann:2005sw,Garcia-Recio:2013gaa,Wu:2010jy,Wu:2010vk,Xiao:2013yca}.
Coupled-channel calculation was considered in Refs.~\cite{Hofmann:2005sw,Garcia-Recio:2013gaa}, but the obtained masses of charm pentaquark were much smaller (less than 4 GeV) than the values observed in LHCb.
Other coupled-channel calculations gave the masses close to the observed ones~\cite{Wu:2010jy,Wu:2010vk,Xiao:2013yca}.
Effect of the direct quark exchange in the hadronic molecule was considered in Ref.~\cite{Wang:2011rga}.
As a compact state, diquark model was analyzed in Ref.~\cite{Maiani:2015vwa}.
QCD sum rules were applied and the mass values close to the observed ones were reported~\cite{Chen:2015moa}.
As other possibities, the cusp effect by a triangle anomaly was discussed~\cite{Liu:2015fea}, and new experimental setup for pion beam was proposed~\cite{Liu:2016dli}.
More references will be found in Ref.~\cite{Chen:2016qju}.

Among many candidates of internal structure, we will consider the compact multiquark state.
We focus on a new possible structure of charm pentaquark with quark configuration $c\bar{c}uds$, which will be denoted by $P_{cs}$, and investigate the mass ordering of $P_{cs}$ for different quantum numbers.

Let us consider the color structure and the light flavor structure in the pentaquark $c\bar{c}qqq$ with $q=u$, $d$ or $s$.
We assume that the pentaquark is a compact quark state, and consider quantum number of $c\bar{c}qqq$ clusters $c\bar{c}$ and $qqq$ separately.
We note that, due to the color singlet condition for hadrons, $c\bar{c}$ and $qqq$ can be not only color singlet but also color octet.
Let us consider the decomposition of the flavor-spin multiplet for $qqq$ in terms of $\mathrm{SU}(6)$ symmetry including $\mathrm{SU}(3)$ flavor symmetry and $\mathrm{SU}(2)$ spin symmetry:
\begin{eqnarray}
 {\bf 6} \times {\bf 6} \times {\bf 6}
= {\bf 20}_{\mathrm{A}} + {\bf 70}_{\mathrm{MA}} + {\bf 70}_{\mathrm{MS}} + {\bf 56}_{\mathrm{S}},
\end{eqnarray}
where the subscripts stand for totally asymmetric (A), mixed asymmetric (MA), mixed symmetric (MS) and totally symmetric (S) cases.
Each multiplet is separated as a sum of flavor and spin,
\begin{eqnarray}
 {\bf 20} &=& ({\bf 8},2) + ({\bf 1},4), \label{eq:fs20} \\
 {\bf 70} &=& ({\bf 8},4) + ({\bf 10},2) + ({\bf 8},2) + ({\bf 1},2), \label{eq:fs70} \\
 {\bf 56} &=& ({\bf 10},4) + ({\bf 8},2), \label{eq:fs56}
\end{eqnarray}
where the first term and second term in the parentheses represent the flavor multiplet and the multiplicity of spin, respectively.

Let us consider the simplest case. 
In the following, we consider all the particles are in $S$-wave, when $c\bar{c}$ is color singlet and $qqq$ is also color singlet.
Then, the light flavor of $qqq$ is given by ${\bf 56}_{\mathrm{S}}$, because the color part of $qqq$ is totally antisymmetric.
Thus, we obtain the well-known multiplet, flavor octet with spin $1/2$ and flavor decuplet with spin $3/2$.

In contrast, the situation is different for the case that $c\bar{c}$ is color octet.
In this case, the color of $qqq$ should be color octet.
According to the decomposition of the color multiplet for three particles,
\begin{eqnarray}
 {\bf 3} \times {\bf 3} \times {\bf 3} = {\bf 1}_{\mathrm{A}} + {\bf 8}_{\mathrm{MA}} + {\bf 8}_{\mathrm{MS}} + {\bf 10}_{\mathrm{S}},
\end{eqnarray}
we have two candidates of color octet, ${\bf 8}_{\mathrm{MA}}$ and ${\bf 8}_{\mathrm{MS}}$, as mixed symmetry state.
Importantly, the combination of the two mixed symmetry states from SU(6) flavor-spin symmetry and SU(3) color symmetry gives the totally antisymmetric state.
It is given by a sum of the tensor product of ${\bf 70}_{\mathrm{MS}}$ and ${\bf 8}_{\mathrm{MA}}$ and the tensor product of ${\bf 70}_{\mathrm{MA}}$ and ${\bf 8}_{\mathrm{MS}}$.
Because of the decomposition of flavor-spin multiplet ${\bf 70}$ in Eq.~(\ref{eq:fs70}), we can consider four flavor-spin multiplets, namely $({\bf 8},4)$, $({\bf 10},2)$, $({\bf 8},2)$ and $({\bf 1},2)$.
In the present study, we will focus on
$({\bf 1},2)$, because this multiplet becomes most stable in the color-spin interaction.

In the literature, there have been studies of internal configurations of charm pentaquark $c\bar{c}qqq$ as a compact state~\cite{Yuan:2012wz,Takeuchi:2016ejt}.
In this reference, the hyperfine splitting was provided by each of color-spin interaction, flavor-spin interaction and instanton-induced interaction, while the confinement potential for quarks was provided by the harmonic oscillator potential.
In a similar idea, we also will use the color-spin interaction and the instanton-induced interaction at short distance, but we adopt the linear  potential as quark confinement potential.
In our case, we include the simultaneous combination of the color-spin interaction and the instanton-induced interaction, and consider the three-body force in the instanton-induced interaction which has not been considered so far.
Furthermore, we investigate the details of the internal spatial structure in $P_{cs}$.

The paper is organized as follows.
In Section 2, we summarize the quark wave function of the charm pentaquark and introduce the setup of the quark model with color-spin interaction, the instanton-induced interaction and quark confinement potential.
We prepare two models. One is given by the color-spin interaction, and another is given by the combination of the color-spin interaction and the instanton-induced interaction.
In Section 3, we perform the variational calculation for mass of $P_{cs}$, and investigate the internal color, spin and spatial structures of the obtained states.
In Section 4, we discuss the possible decay modes of $P_{cs}$.
The final section is devoted to our conclusion.

\section{Quark model}

\subsection{Wave function of charm pentaquark}

\begin{figure}[tbp]
  \centering
  \includegraphics[keepaspectratio,scale=0.15]{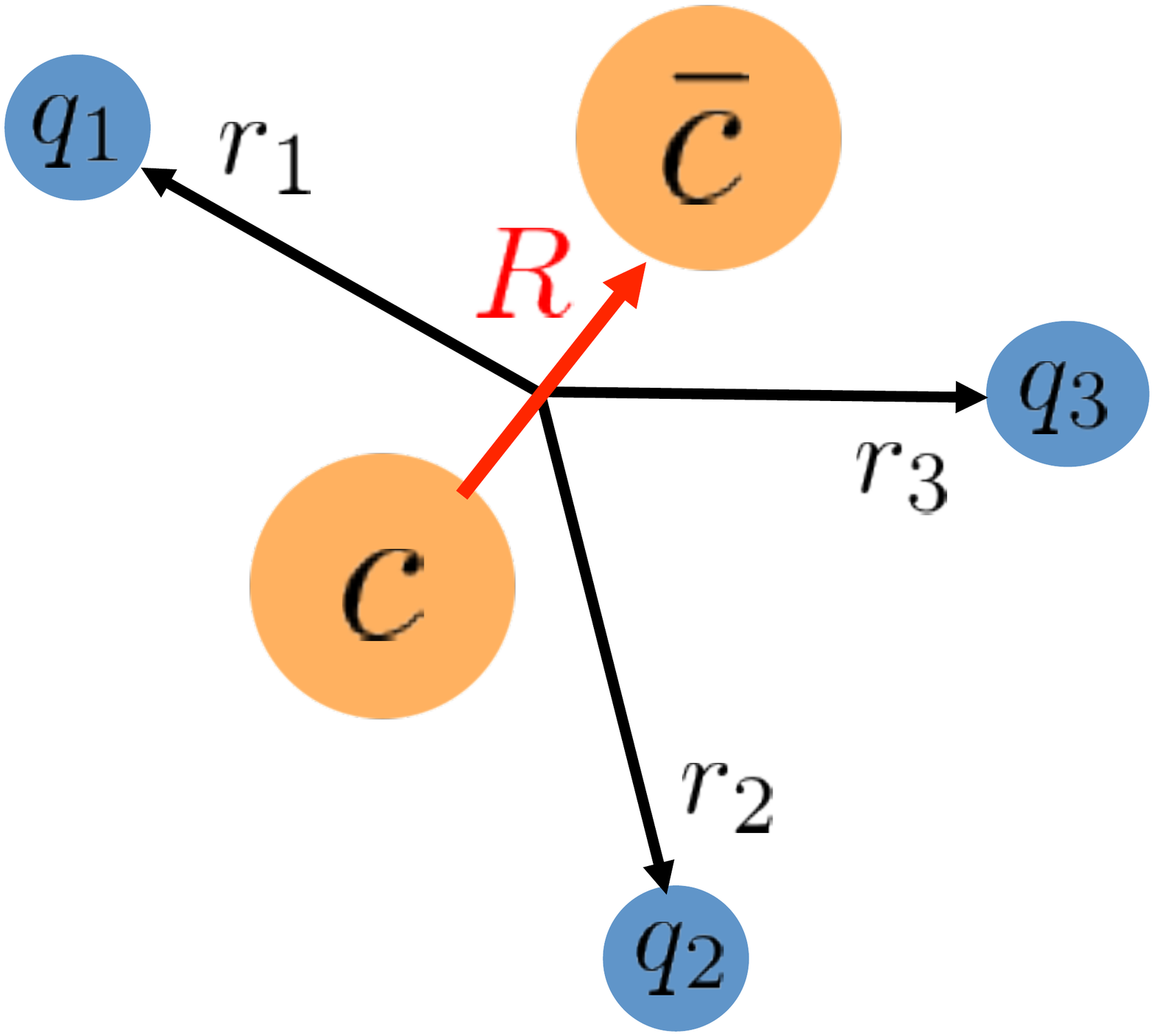}
\caption{The coordinate of $\vec{R}$ and $\vec{r}_{i}$ ($i=1,2,3$).}
  \label{fig:cooridinate}
\end{figure}

For $P_{cs}$ ($c\bar{c}uds$), we consider that the total wave function is given by a product of the spatial part ($\phi$), the spin and color part of $c\bar{c}$ ($\psi_{c\bar{c}}^{s,c}$), and the spin, color and flavor part ($\psi_{uds}^{s,c,f}$):
\begin{eqnarray}
\psi = \phi(\vector{R},\vector{r}_1,\vector{r}_2,\vector{r}_3) \psi_{c\bar{c}}^{s,c} \psi_{uds}^{s,c,f}.
\end{eqnarray}
The spatial part $\phi$ depends on the variables $\vector{R}$ and $\vector{r}_{i}$ ($i=1,2,3$) (Fig.~\ref{fig:cooridinate}).
Here $\vector{R}$ is the position vector from the $c$ quark to the $\bar{c}$ quark, and $\vector{r}_{i}$ are the vectors for light quarks $i=1,2,3$.
We assume for simplicity that the internal angular momenta are $S$-wave because we focus on the ground states.

It is known that the Jacobi coordinates are very useful to solve many-body problems in general.
In the present discussion, however, we simplify the situation in the following way.
We assume that the $c$ and $\bar{c}$ quarks are sufficiently heavy, and that the midpoint of $c$ and $\bar{c}$, i.e. $\vector{R}/2$, represents the center-of-mass of $c\bar{c}qqq$ system.
In this limiting case, we can assign the original points of the vectors $\vector{r}_{i}$ to be the center-of-mass of the system.
We notice that in this treatment the motion of the $c\bar{c}$ (or $uds$) cluster to the total system is neglected.
Nevertheless, we expect that this would be a reasonable approximation as long as the mass of charm quark is much larger than those of light quarks.

As for the spatial wave function,
we here consider only compact systems of five quarks and assume the Gaussian type with extension parameters $a$ for $\vector{R}$ and $b$ for $\vector{r}_{i}$.
We use the common value $b$ for $\vector{r}_{1}$, $\vector{r}_{2}$, $\vector{r}_{3}$, because the wave function of the light quarks will be distributed uniformly in space.
In fact, as will be discussed later, the stability of the pentaquark considered here seems irrelevant to the diquark correlation between two light quarks, but rather sensitive to the $c\bar{c}$ correlations.
In this sense, we may justify to treat the common variational parameter $b$.

With the simplifications stated above, we assume the spatial part of the wave function as
\begin{eqnarray}
\phi(\vector{R},\vector{r}_1,\vector{r}_2,\vector{r}_3)=\frac{1}{(2\pi a^2)^
\frac{3}{4}}\frac{1}{(\pi b^2)^\frac{9}
{4}}\exp\left(-\frac{|\vector{R}|^2}{4a^2}-\frac{|\vector{r}_1|^2+|\vector{r}_2|^2+|\vector{r}_3|^2}{2b^2}\right),
\label{eq:kukan}
\end{eqnarray}
which is normalized by integrating over the space.
The values of $a$ and $b$ will be determined by variational calculation.
Note that all the orbital angular momenta are zero.

\begin{table}[tb]
\begin{center}
	\caption{Combinations of internal color states of $P_{cs}$ with isospin $I=0$ and spin-parity $J^{P}$. They are denoted by $P_{cs\mathbf{8}}$ ($1/2^{-}$), $P'_{cs\mathbf{8}}$ ($1/2^{-}$), $P^{\ast}_{cs{\mathbf{8}}}$ ($3/2^{-}$) for each $J^{P}$ in color octet type ($\mathbf{8}$) for component $c\bar{c}$ (or $uds$), and by $P_{cs\mathbf{1}}$ ($1/2^{-}$), $P'_{cs\mathbf{1}}$ ($1/2^{-}$), $P^\ast_{cs\mathbf{1}}$ ($3/2^{-}$) for each $J^{P}$ in color singlet type ($\mathbf{1}$) for component $c\bar{c}$ (or $uds$). Notice that the spin combination of $c\bar{c}$ is different for $P_{cs\mathbf{8}}$ and $P'_{cs\mathbf{8}}$ ($P_{cs\mathbf{1}}$ and $P'_{cs\mathbf{1}}$): spin 0 for the former and spin 1 for the latter.}
	\begin{tabular}{|c||c|c|c|c|c|c||c|c|c|c|c|c|}
	        \hline
		\multicolumn{1}{|c||}{$(I,J^P)$}& \multicolumn{6}{c||}{octet type ($\mathbf{8}$)} & \multicolumn{6}{c|}{singlet type ($\mathbf{1}$)} \\
		\cline{2-13}
		 & & component&color&spin&flavor&isospin& & component& color&spin&flavor&isospin
		\\ \hline\hline
	$(0,1/2^-)$& $P_{cs\mathbf{8}}$ &$\ccbar$&$\hati$&0& --- & --- & $P_{cs\mathbf{1}}$ & $\ccbar$ &$\iti$&0& --- & --- \\ 
	\cline{3-7} \cline{9-13}
	& &$\uds$&$\hati$&1/2&$\iti$&0& & $\uds$ & $\iti$&1/2&  $\hati$&0\\ \hline
	\hline
	$(0,1/2^-)$& $P'_{cs\mathbf{8}}$ &$\ccbar$&$\hati$&1& --- & --- & $P'_{cs\mathbf{1}}$ & $\ccbar$ &$\iti$&1& --- & --- \\ 
	\cline{3-7} \cline{9-13}
          & & $\uds$&$\hati$&1/2&$\iti$&0& & $\uds$ & $\iti$&1/2&$\hati$&0\\ \hline
          \hline
	$(0,3/2^-)$& $P^{\ast}_{cs\mathbf{8}}$ & $\ccbar$&$\hati$&1& --- & --- & $P^{\ast}_{cs\mathbf{1}}$ & $\ccbar$ &$\iti$&1& --- & --- \\ 
	\cline{3-7} \cline{9-13}
         & &$\uds$&$\hati$&1/2&$\iti$&0& & $\uds$ &$\iti$&1/2&$\hati$&0 \\
         \hline
\end{tabular}
\label{table:internal}
\end{center}
\end{table}

As for the spin-color part of $c\bar{c}$ ($\psi_{c\bar{c}}^{s,c}$) and spin-color-flavor part of $uds$ ($\psi_{uds}^{s,c,f}$), we consider several combinations of quantum numbers as summarized in Table~\ref{table:internal}.
As for spin, we consider the cases where the spin of $c\bar{c}$ is either 0 or 1, and the spin of $uds$ is 1/2.
Then, the total spin and parity of the the charm pentaquark is $J^{P}=1/2^{-}$ with $c\bar{c}$ spin 0 or 1, and $J^{P}=3/2^{-}$ with $c\bar{c}$ spin 1.
We notice that $c\bar{c}$ with spin 0 and $c\bar{c}$ with spin 1 should be regarded as the independent states which are not mixed with each other.
This observation is supported by the fact that the spin of charm quark is conserved in the heavy quark mass limit, as known in the heavy quark effective theory.
In reality, however, there is a small correction term which breaks the heavy quark spin symmetry with an order of $1/m_{c}$, and it induces the mixing of $c\bar{c}$ spin 0 and $c\bar{c}$ spin 1.

First, we consider $\psi_{csc\bar{c}}^{s,c}$.
This is composed of the spin part ($\chi^{s}_{\cc}$) and the color part ($\psi^{c}_{\cc}$):
\begin{eqnarray}
  \psi_{\ccbar}^{s,c} = \psi^{c}_{\cc} \, \chi^{s}_{\cc},
  \label{eq:ccbar}
\end{eqnarray}
with $c=\mathbf{1}$ for color singlet and $c=\mathbf{8}$ for color octet, and $s=0$ for spin singlet and $s=1$ for spin triplet.

Second, as for $\psi_{uds}^{s,c,f}$, we consider the following combinations of color part ($\psi_{uds}^{c}$), spin part ($\chi_{uds}^{s}$), and flavor part ($\psi_{uds}^{f}$).
In the case of three particles $uds$, we have to pay a special attention to the antisymmetriation of the wave functions.
Because all the internal angular momenta are $S$-wave, the combination of color, spin and flavor of $uds$ should be antisymmetric.
We consider the color octet case and the color singlet case for $uds$.
Let us first consider the case of flavor singlet $f=\mathbf{1}$.
In this case, the combination of color and spin needs to be totally symmetric, because the flavor part is totally antisymmetric.
Then the allowed combination of the color and spin is
\begin{eqnarray}
 \frac{1}{\sqrt{2}}
 \bigl(
  \psi_{uds}^{c=\mathbf{8}_{\lambda}} \chi_{uds}^{s=1/2_{\lambda}}
  + \psi_{uds}^{c=\mathbf{8}_{\rho}} \chi_{uds}^{s=1/2_{\rho}}
 \bigr),
\end{eqnarray}
where the subscript $\lambda$ ($\rho$) in $\mathbf{8}_{\lambda}$ ($\mathbf{8}_{\rho}$) and $1/2_{\lambda}$ ($1/2_{\rho}$) means that the first two light quarks are symmetric (antisymmetric) under exchange of the two light quarks.
The product of $\rho$ state and $\lambda$ state makes the totally symmetric state under exchange of any two light quarks~\footnote{As another combination, we may consider the $\mathbf{10}$ representation for color and spin $3/2$, namely $P_{cs}^{c=\mathbf{10}}\chi_{uds}^{s=3/2}$. However, the $\mathbf{10}$ representation of color is not allowed for $uds$ in $c\bar{c}uds$, because $uds$ should be $\mathbf{1}$ or $\mathbf{8}$ in accordance to the possible color representation of $c\bar{c}$.}.
Then, we have the $uds$ wave function for light flavor singlet, $f=\mathbf{1}$:
\begin{eqnarray}
\psi_{uds}^{s=1/2,c=\hati,f=\iti,I=0}
=\frac{1}{\sqrt{2}}
\Bigl( \psi^{c=\hati_{\lambda}}_{uds}\chi^{s=1/2_{\lambda}}_{uds}+\psi^{c=\hati_{\rho}}_{uds}\chi^{s=1/2_{\rho}}_{uds} \Bigr) \psi^{f=\iti,I=0}_{uds},
\label{eq:uds1}
\end{eqnarray}
where $\psi^{f=\iti,I=0}_{uds}$ is the flavor singlet wave function.
We add an upper script $I=0$, because we will consider isospin singlet $I=0$ only.
Second, we consider the light flavor octet, $f=\mathbf{8}$.
In this case, by combining the light flavor and the spin for light quarks, we may consider the totally symmetric state for flavor and spin,
\begin{eqnarray}
  \frac{1}{\sqrt{2}} \Bigl( \chi^{s=1/2_{\lambda}}_{uds}\psi^{f=\hati_{\lambda},I=0}_{uds}+\chi^{s=1/2_{\rho}}_{uds}\psi^{f=\hati_{\rho},I=0}_{uds} \Bigr),
\end{eqnarray}
where $\lambda$ ($\rho$) is the same notation as before.
We add $I=0$ for flavor wave function, because we will consider $I=0$ only.
The color part should be totally antisymmetric, $\psi^{c=\iti}_{uds}$.
Hence we obtain the $uds$ wave function for flavor octet, $f=\mathbf{8}$:
\begin{eqnarray}
\psi_{uds}^{s=1/2,c=\iti,f=\hati,I=0}
=\psi^{c=\iti}_{uds}
\frac{1}{\sqrt{2}} \Bigl( \chi^{s=1/2_{\lambda}}_{uds}\psi^{f=\hati_{\lambda},I=0}_{uds}+\chi^{s=1/2_{\rho}}_{uds}\psi^{f=\hati_{\rho},I=0}_{uds} \Bigr).
\label{eq:uds2}
\end{eqnarray}

By combining $P_{csc\bar{c}}^{s,c}$ in Eq.~(\ref{eq:ccbar}) and $P_{cs}^{s,c,f}$ in Eqs.~(\ref{eq:uds1}) and (\ref{eq:uds2}), we will have four states in $(I,J^{P})=(0,1/2^{-})$ and two states in $(0,3/2^{-})$.
Their explicit forms are
\begin{eqnarray}
\po(\vector{R},\vector{r}_1,\vector{r}_2,\vector{r}_3)
&=& \phi_{\mathbf{8}}(\vector{R},\vector{r}_1,\vector{r}_2,\vector{r}_3)
\Bigl[ \psi_{\ccbar}^{s=0,c=8} \otimes  \psi_{uds}^{s=1/2,c=8,f=1,I=0} \Bigr]^{s=1/2}, \nonumber \\
\ps(\vector{R},\vector{r}_1,\vector{r}_2,\vector{r}_3)
&=&\phi_{\mathbf{1}}(\vector{R},\vector{r}_1,\vector{r}_2,\vector{r}_3)
\Bigl[ \psi_{\ccbar}^{s=0,c=1} \otimes \psi_{uds}^{s=1/2,c=1,f=8,I=0} \Bigr]^{s=1/2},
\label{eq:pentasix1}
\end{eqnarray}
for $c\bar{c}$ spin 0 and $(I,J^{P})=(0,1/2^{-})$,
\begin{eqnarray}
\pop(\vector{R},\vector{r}_1,\vector{r}_2,\vector{r}_3)
&=&\phi'_{\mathbf{8}}(\vector{R},\vector{r}_1,\vector{r}_2,\vector{r}_3)
\Bigl[ \psi_{\ccbar}^{s=1,c=8} \otimes \psi_{uds}^{s=1/2,c=8,f=1,I=0} \Bigr]^{s=1/2}, \nonumber \\
\psp(\vector{R},\vector{r}_1,\vector{r}_2,\vector{r}_3)
&=&\phi'_{\mathbf{1}}(\vector{R},\vector{r}_1,\vector{r}_2,\vector{r}_3)
\Bigl[ \psi_{\ccbar}^{s=1,c=1} \otimes \psi_{uds}^{s=1/2,c=1,f=8,I=0} \Bigr]^{s=1/2},
\label{eq:pentasix2}
\end{eqnarray}
for $c\bar{c}$ spin 1 and $(I,J^{P})=(0,1/2^{-})$,
\begin{eqnarray}
\poa(\vector{R},\vector{r}_1,\vector{r}_2,\vector{r}_3)
&=&\phi^{\ast}_{\mathbf{8}}(\vector{R},\vector{r}_1,\vector{r}_2,\vector{r}_3)
\Bigl[ \psi_{\ccbar}^{s=1,c=8} \otimes \psi_{uds}^{s=1/2,c=8,f=1,I=0} \Bigr]^{s=3/2}, \nonumber \\
\psa(\vector{R},\vector{r}_1,\vector{r}_2,\vector{r}_3)
&=&\phi^{\ast}_{\mathbf{1}}(\vector{R},\vector{r}_1,\vector{r}_2,\vector{r}_3)
\Bigl[ \psi_{\ccbar}^{s=1,c=1} \otimes \psi_{uds}^{s=1/2,c=1,f=8,I=0} \Bigr]^{s=3/2},
\label{eq:pentasix3}
\end{eqnarray}
for $c\bar{c}$ spin 1 and $(I,J^{P})=(0,3/2^{-})$,
where the subscripts $\mathbf{8}$ and $\mathbf{1}$ indicate that the color representation of the components, $c\bar{c}$ and $uds$,
and the square brackets indicate the composition of total spin.
Notice that the spatial wave functions are different for each color and spin, as denoted by $\phi_{\mathbf{8},\mathbf{1}}$, $\phi'_{\mathbf{8},\mathbf{1}}$ and $\phi^{\ast}_{\mathbf{8},\mathbf{1}}$.

\subsection{Model A: hamiltonian {\it without} instanton interaction}

We consider the Hamiltonian for $c\bar{c}uds$.
It is given as sum of the kinetic term ($K$), the color-Coulomb term ($V_{\mathrm{Coulomb}}$), the color-magnetic interaction (CMI) term ($V_{\mathrm{CMI}}$) and the confinement term ($V_{\mathrm{conf}}$):
\begin{eqnarray}
H_{A} = K+V_{\mathrm{Coulomb}}+V_{\mathrm{CMI}}+V_{\mathrm{conf}},
\label{eq:hamiltonian_A}
\end{eqnarray}
where each term is given by
\begin{eqnarray}
K&=& - \frac{\vector{\nabla}^{2}_{R}}{2\mu_{c\bar{c}}} - \frac{\vector{\nabla}^{2}_{1}}{2m_1}-\frac{\vector{\nabla}^{2}_{3}}{2m_2}-\frac{\vector{\nabla}^{2}_{3}}{2m_3}, \\
	\label{eq:kin}
V_{\mathrm{Coulomb}}&=&\sum_{i<j}\frac{\alpha_s}{4 r_{ij}} \vector{\lambda}_i \cdot \vector{\lambda}_j, 
   \label{eq:Coulomb} \\
V_{\mathrm{CMI}}&=&-\frac{\alpha_s}{4}\sum_{i<j}\frac{\pi}{m_im_j}
	\vector{\lambda}_i \cdot \vector{\lambda}_j\left(1+\frac{2}{3}\vector{\sigma}_i \cdot \vector{\sigma}_j \right) \delta^{(3)}(r_{ij}), 
	\label{eq:CMI} \\
\conf&=& -\sigma \, \sum_{i<j} \vector{\lambda}_i \cdot \vector{\lambda}_j \, r_{ij},
\label{eq:conf}
\end{eqnarray}
where we define $\vector{\nabla}_{R} = {\partial}/{\partial \vector{R}}$ and $\vector{\nabla}_{k} = {\partial}/{\partial \vector{r}_{k}}$ ($k=1,2,3$), $\vector{\lambda}_{i}$ and $\vector{\sigma}_{i}$ the Gell-Mann matrices for color and the Pauli matrices for spin for quarks $i=c$, $\bar{c}$, $q_{1}$, $q_{2}$ and $q_{3}$, $r_{ij}=|\vector{r}_{i}-\vector{r}_{j}|$ the distance between the quark $i$ and $j$.
The hadron mass is given by the sum of the expectation value of $\langle H_{A} \rangle$ and a constant term $C$: $E=\langle H_{A} \rangle+C$.
As parameters we use $\alpha_{s}$ for the coupling constant in the Coulomb potential and the CMI potential, $\mu_{c\bar{c}}=m_{c}/2$ with charm quark mass $m_{c}$ and $m_{k}$ the mass for light quark $k=1,2,3$, and $\sigma$ the string tension of the linear confinement potential.
As for $\alpha_{s}$ and $\sigma$,
we use different values for light-light quark pairs and for light-heavy and heavy-heavy quark pairs.
The parameters in the former are denoted by $\alpha_{s1}$ and $\sigma_{1}$, and the ones for the latter are by $\alpha_{s2}$ and $\sigma_{2}$.
We use the one-third of the nucleon mass for $m_{u}=m_{d}$, and $m_{s}$ is from the mass ratio $m_{u}/m_{s}=0.6$ so
that they reproduce the masses of the light ground-state baryons, as summarized in Table~\ref{table:baryons}.
The constant term $C_{\Lambda}$ is adjusted to the $\Lambda$ baryon.
In the heavy sector, the values of $m_{c}$ for c quark mass, $\alpha_{s2}$ for the coupling constant between two heavy quarks (or a heavy quark and a light quark), $\sigma_{2}$ for the string tension between two heavy quarks (or a heavy quark and a light quark), and the constant $C_{\eta_{c}}$ for $\eta_{c}$ are taken from Ref.~\cite{Barnes:2005pb}, which reproduce the masses of $\eta_{c}$ and $J/\psi$.

\begin{table}[tbp]
	\begin{center}
	\caption{Parameter sets of the model A and the model B. We use the notation in the model B as L-L: pair of a light quark and a light quark, H-H: pair of a heavy quark and heavy quark pair, H-L: pair of a heavy quark and a light quark.}
		\begin{tabular}{|c|c|c|}
			\hline
			                      & model A & model B \\
			\hline \hline
			$m_u$ [MeV] & 313 & 313  \\ \hline
			$m_s$ [MeV] & 521.7 & 521.7 \\ \hline
			$m_c$ [MeV] & 1497.4 & 1497.4 \\ \hline
			$\alpha_{s1}$ & 0.769 & 0.715 \\ \hline
			$\alpha_{s2}$ & 0.5461 & 0.5461 \\ \hline
			$\sigma_1$ [MeV/fm] & 178 & 178 \\ \hline
			$\sigma_2$ [MeV/fm] & 135.63 & 135.63 \\ \hline
			$C_\Lambda$ [MeV] & -1130 & -1470 \\ \hline
			$C_{\eta_c}$ [MeV] & -61 & -61 \\ \hline
			$U_0^{(2)}$ & --- &-1.331 \\ \hline
			$V_0$ [MeV$^{-5}$] & --- & $5.271\times10^{-13}$ \\ \hline
			$p$ (L-L) & --- & 0.4 \\ \hline
			$p$ (H-H,H-L) & --- & 0 \\
			\hline
		\end{tabular}
	\label{table:para1}
	\end{center}
\end{table}

\begin{table}[tbp]
	\begin{center}
		\caption{Masses of normal baryons with up, down and strangeness in the models A and B. Units are in MeV.}
		\begin{tabular}{|c||c|c|c|}\hline
			baryon &  model A & model B & experiments~\cite{Olive:2016xmw} \\ \hline\hline
			$N(1/2^+)$ & 1048 & 1019 & 939 \\ \hline
			$\Delta(3/2^+)$ &  1247 & 1220 & 1232 \\ \hline
			$\Lambda(1/2^+)$  & 1116 & 1116 & 1116 \\ \hline
			$\Sigma(1/2^+)$  & 1193 & 1193 & 1193 \\ \hline
			$\Sigma^\ast(3/2^+)$  & 1330 & 1327 &1385 \\ \hline
		\end{tabular}
		\label{table:baryons}
	\end{center}
\end{table}

\begin{table}[tbp]
\begin{center}
\caption{The expectation values of $\vector{\lambda}_{i} \cdot \vector{\lambda}_{j} \, \vector{\sigma}_{i} \cdot \vector{\sigma}_{j}$ for a pair of quark $i$ and $j$. Notice $\vector{\sigma}_{i} \cdot \vector{\sigma}_{j}=-3$ for spin singlet ($s=0$) and $1$ for spin triplet ($s=1$).}
\begin{tabular}{|c|c|c|c|c|c|}
\hline
 \multicolumn{2}{|c}{} & \multicolumn{4}{|c|}{color} \\
\cline{3-6}
 \multicolumn{2}{|c|}{} & $c=\mathbf{1}$ & $c=\mathbf{8}$ & $c=\bar{\mathbf{3}}$ & $c=\mathbf{6}$ \\
\hline
spin & $s=0$ & $+16$ & $-2$ & $+8$ & $-4$ \\
       & $s=1$ & $-\frac{16}{3}$ & $+\frac{2}{3}$ & $-\frac{8}{3}$ & $+\frac{4}{3}$ \\
\hline
\end{tabular}
\end{center}
\label{table:operators}
\end{table}%

\subsection{Model B: hamiltonian {\it with} instanton interaction}

In the model A, we have considered the one-gluon exchange potential at short distance.
However, there can be additional interaction which originates from the instanton.
The instanton is responsible for the $\mathrm{U}(1)_{\mathrm{A}}$ breaking in QCD vacuum, and can be seen in several mass spectrum of hadrons.
One of the most prominent effects is seen in $\eta'$ mass, whose mass is much larger than the other Nambu-Goldstone bosons ($\pi$, $\eta$, $K$).
Another example can be seen in H-dibaryons ($uuddss$)~\cite{Takeuchi:1990qj}.
The instanton couples to massless quarks strongly through zero modes, and generates a six-quark vertex given by a three-body force in the flavor singlet channel.
Indeed, the instanton has the property that there exists a zero-energy bound state of massless fermion around the instanton~\cite{tHooft:1976snw}.
In our case, $uds$ in the charm pentaquark $c\bar{c}uds$ can be flavor singlet (cf.~Table~\ref{table:internal}), and hence the instanton may play an interesting role.

Let us summarize briefly the properties of the instanton.
The instanton configuration is given by
\begin{eqnarray}
 (G_{\mu\nu}^a)^2=\frac{192\rho^4}{(x^2+\rho^2)^4},
\end{eqnarray}
as the classical solution of QCD in four-dimensional Euclidean space.
The parameter $\rho$ is the instanton size.
It is estimated as about $0.3$ fm in the instanton liquid model~\cite{Shuryak:1984nq}.
This size is smaller than the typical hadron size, $1/\Lambda_{\mathrm{QCD}} \sim 1$ fm for $\Lambda_{\mathrm{QCD}} \simeq 200$ MeV.
Therefore, it is possible to regard the instanton as a point-like object and the effective interaction between quarks via instanton can be represented by a point-like interaction.

\begin{figure}[tb]
\begin{center}
  \begin{minipage}[b]{0.49\linewidth}
    \centering
    \includegraphics[keepaspectratio, angle=270, scale=0.18]{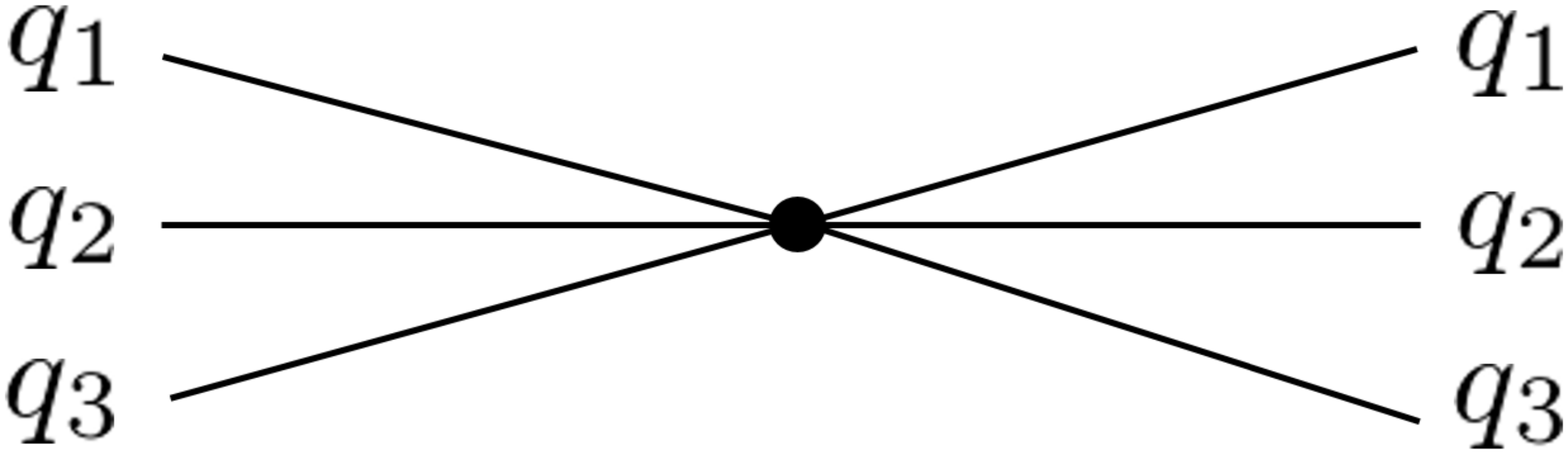}
  \end{minipage}
  \begin{minipage}[b]{0.49\linewidth}
    \centering
    \includegraphics[keepaspectratio, angle=270, scale=0.18]{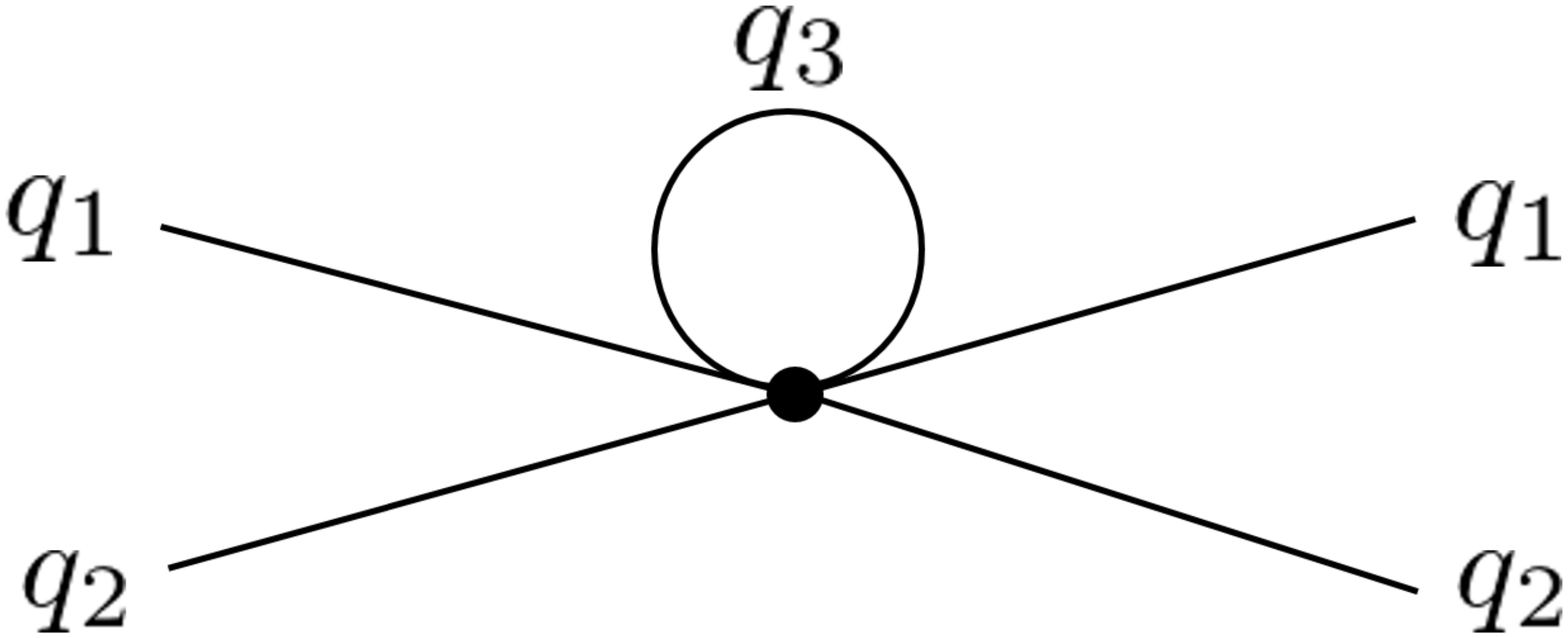}
  \end{minipage}
  \caption{The instanton-induced interaction for six-quark vertex (three-body force) (left) and four-quark vertex (two-body force) (right).}
  \label{fig:instanton32}
  \end{center}\vspace{-5mm}
\end{figure}

The non-relativistic form of Hamiltonian of the instanton-induced interaction for quarks via instanton can be given by
\begin{eqnarray}
H^{(3)} &=& -\mathcal{L}^{(3)}_{\mathrm{eff}} \nonumber \\
&=&V_0\bar{\psi}_{R}(1)\bar{\psi}_{R}
(2)
\bar{\psi}_{R}
(3)\frac{189}{40}\mathcal{A}_3^f \Bigl( 1-\frac{1}{7}\sum^3_{i<j}\vector{\sigma}_i \cdot \vector{\sigma}_j \Bigr)
\psi_{L}(3)\psi_{L}(2)\psi_{L}(1) + \mathrm{h.s.}, 
\label{eq:instanton3}
\end{eqnarray}
with $\psi_{R}(i)=\frac{1}{2}(1+\gamma_{5})\psi(i)$ and $\psi_{L}(i)=\frac{1}{2}(1-\gamma_{5})\psi(i)$ for light quark $i=1,2,3$~\cite{Takeuchi:1990qj,Takeuchi:1992sg}.
This is a six-quark vertex, namely the three-body force (Fig.~\ref{fig:instanton32}).
The three flavors of quarks should be different, because the projection operator for ansisymmetrization of light flavor, $\mathcal{A}_3^f$, is introduced to pickup the flavor singlet component.
The parameter $V_{0}$ is the coupling constant, whose value can be determined phenomenologically.
It is noted that the second term in the r.h.s., the hermitian conjugate to the first term, represents the contribution from the anti-instanton.

The three-body force of the instanton-induced interaction can be transformed to the two-body force.
This is indeed accomplished by closing one pair of quarks ($q_{3}$ in Fig.~\ref{fig:instanton32}) with a quark condensate $\langle \bar{\psi}\psi \rangle$, and the obtained interaction is given by
\begin{eqnarray}
H^{(2)}&=&-\mathcal{L}^{(2)}_{\mathrm{eff}} \nonumber \\
&=&V_0^{(2)}(1,2) \bar{\psi}_{R}(1)\bar{\psi}_{R}(2) \frac{15}{8}\mathcal{A}_2^f \Bigl (1-\frac{1}{5} \vector{\sigma}_1 \cdot \vector{\sigma}_2 \Bigr)
\psi_{L}(2)\psi_{L}(1)+\mathrm{h.c.},
\label{inslag2}
\end{eqnarray}
as the effective interaction for $q_{1}$ and $q_{2}$.
The effective coupling constant $V_0^{(2)}(1,2)$ is a product of $V_{0}$ in the three-body force and the loop of $q_{3}$, namely the chiral condensate of $q_{3}$, and the explicit form is given by
\begin{eqnarray}
V_0^{(2)}(1,2)
&=&\frac{1}{2}V_0 \Bigl( \braket{\bar{\psi}\psi}-Km_3^{(c)} \Bigl)
\nonumber \\
&=& \frac{1}{2} V_{0} K m_{3}.
\end{eqnarray}
It should be noted that the current mass $m_{3}^{(c)}$ of the quark $q_{3}$ is included also in the second term in the parenheses.
$K$ is the coefficient for connecting the constituent mass of quark $q_{3}$, $m_{3}$, and the current mass $m_{3}^{(c)}$ and the chiral condensate $\langle \bar{\psi}\psi \rangle$:
\begin{eqnarray}
m_3 \equiv m_3^{(c)}-\frac{1}{K}\braket{\bar{\psi}\psi}.
\end{eqnarray}
Then, the effective coupling constant can be eventually represented as
\begin{eqnarray}
V_0^{(1,2)}=-\frac{1}{2}V_0Km_3.
\end{eqnarray}
The value of $K$ should be in principle dependent on quark flavor.
Nevertheless, we assume the SU(3) flavor symmetry, and adopt the value of $K$ estimated in the $u$ quark sector.
By using the current mass $m_{u}^{(c)}=2.2$ MeV, the constituent mass $m_{u}=313$ MeV, and the chiral condensate $\langle \bar{\psi}\psi \rangle=-(250 \hbox{ MeV})^{3}$, we obtain the value
\begin{eqnarray}
K=-\frac{\braket{\bar{\psi}\psi}}{m_u-m_u^{(c)}}=-5.027 \hspace{0.5em} [\mathrm{MeV}^2].
\end{eqnarray}
Finally, by defining the effective coupling by
\begin{eqnarray}
U_0^{(2)}=-\frac{1}{2}V_0Km_u^2m_s,
\label{eq:UVkankei}
\end{eqnarray}
we obtain the effective two-body interaction
\begin{eqnarray}
H^{(2)}=-\mathcal{L}_\mathrm{eff}^{(2)}=U_0^{(2)}\sum_{i<j}\frac{1}{m_im_j}
\bar{\psi}_{R}(1)\bar{\psi}_{R}(2)
\mathcal{A}_2^f \Bigl( 1-\frac{1}{5}\vector{\sigma}_i \cdot \vector{\sigma}_j \Bigr)
\psi_{L}(2)\psi_{L}(1)+\mathrm{h.c.},
\label{eq:instanton2}
\end{eqnarray}
which is much compactly represented in the form that the flavor dependence appears only in $1/(m_{i}m_{j})$.

From the above results for the three-body force and the two-body force, we derive the effective potentials~\cite{Takeuchi:1992sg},
\begin{eqnarray}
\iiid&=&U_0^{(2)}\frac{15}{8}\sum_{i<j} \mathcal{A}_2^f \frac{1}{m_im_j} \Bigl( 1-\frac{1}{5} \vector{\sigma}_i \cdot \vector{\sigma}_j \Bigr)\delta^{(3)}(r_{ij}),
\label{eq:III2_QM}
\end{eqnarray}
and
\begin{eqnarray}
\iiit&=&V_0\frac{189}{40}\sum_{(ijk)} \mathcal{A}_3^f \Bigl( 1-\frac{1}{7}\bigl(\vector{\sigma}_i \cdot \vector{\sigma}_j+\vector{\sigma}_j \cdot \vector{\sigma}_k+\vector{\sigma}_k \cdot \vector{\sigma}_i \bigr) \Bigr)\delta^{(3)}(r_{ij})\delta^{(3)}(r_{jk}),
\label{eq:III3_QM}
\end{eqnarray}
where the spatial dependence between two quarks (three quarks) are represented by the delta-type potentials, $\delta^{(3)}(r)$ with a distance between two quarks $r$.
As we use the variational method with a single Gaussian extension parameter,
we do not smear the delta function in this study.
For complete solutions, we need to smear the delta according to the size of the Fermion zero modes around the instanton.
$\mathcal{A}_2^f$ and $\mathcal{A}_3^f$ are the projection operators to pickup anti-symmetric representation for two-quark $i,j$ and three-quark $i,j,k$, respectively.

It is interesting to notice that the two-body potential, $\iiid$, has the factor $1/m_{i}m_{j}$ and the spin dependence, and hence that $\iiid$ resembles the spin-dependent part of the one-gluon exchange potential, Eq.~(\ref{eq:CMI}).
In this sense, it leaves some ambiguity in phenomenology about whether the spin-dependent interaction is supplied by the one-gluon exchange or by the instanton-induced interaction.
We here introduce a new parameter $p$ to control the contributions from the one-gluon exchange and the instanton-induced interaction in the Hamiltonian:
\begin{eqnarray}
H&=&K+(1-p)\bigl( V_{\mathrm{Coulomb}}+V_{\mathrm{CMI}} \bigr)_{\mathrm{LL}}+p \bigl( \iiid+\iiit \bigr)
\nonumber \\
&& +\bigl( V_{\mathrm{Coulomb}}+V_{\mathrm{CMI}} \bigr)_{\mathrm{HL}}+\bigl( V_{\mathrm{Coulomb}}+V_{\mathrm{CMI}} \bigr)_{\mathrm{HH}}+V_{\mathrm{conf}},
\label{eq:hamil2}
\end{eqnarray}
where the subscripts $\mathrm{LL}$, $\mathrm{HL}$ and $\mathrm{HH}$ indicate the operated pairs of two quarks, light-light quarks ($\mathrm{LL}$), heavy-light quarks ($\mathrm{HL}$) and heavy-heavy quarks ($\mathrm{HH}$).
The light baryon spectroscopy can not fix this value because the total strength of the spin dependent interaction is
independent of $p$.
On the other hand, we can determine $p$ phenomenologically in the light meson sector so that the $\eta'$ mass is reproduced,
giving $p=0.4$.
Note that $p$ affects only the short range interaction among light quarks ($\mathrm{LL}$). The confinement potential $V_{\mathrm{conf}}$ is independent of $p$.
The interactions between heavy-light quarks ($\mathrm{HL}$) and heavy-heavy quarks ($\mathrm{HH}$) are not affected by the instanton-induced interaction, because this interaction acts only on light quarks.
We notice also that the three-body force $\iiit$ is also weighted by $p$.
It is clear that the one-gluon exchange (instanton-induced interaction) is recovered for $p=0$ ($p=1$).

The new parameters $U_{0}^{(2)}$, $V_{0}$ and $p$ in the instanton-induced interaction as well as the parameters in the one-gluon exchange are summarized in Table~\ref{table:para1}.
The parameters for heavy quarks, $m_{c}$, $\alpha_{s2}$, $\sigma_{2}$ and $C_{\eta_{c}}$ are the same as those in the model A.
As for the light quark sector, $m_{u}$, $m_{s}$ and $\sigma_{1}$ are the same also, because they should not depend on the details of the interaction at short distance.
The parameters in the one-gluon exchange and the instanton-induced interaction, $\alpha_{s1}$, $U_{0}^{(2)}$, are determined by the mass splitting between $N$ and $\Delta$ baryons.
It is useful to adopt the relation
\begin{eqnarray}
M_\Delta-M_N=\frac{2\sqrt{2}}{3\sqrt{\pi}}\frac{\alpha_s}{m_u^2b^3}=-
\frac{9\sqrt{2}}
{16\pi^\frac{3}{2}}\frac{U_0^{(2)}}{m_u^2b^3},
\end{eqnarray}
for a single Gaussian wave function with the size parameter $b$ (cf.~Eq.~(\ref{eq:kukan})).
The value of $V_{0}$ is determined from $U_{0}^{(2)}$ by Eq.~(\ref{eq:UVkankei}).
The value of $C_{\Lambda}$ is determined to reproduce the mass of the $\Lambda$ baryon.
The fraction $p=0.4$ is determined by the mass splitting $\eta$-$\eta'$ relevant to $\mathrm{U}(1)_{\mathrm{A}}$ breaking~\cite{Takeuchi:1990qj,Takeuchi:1992sg}.

\section{Numerical results}

\subsection{Variational calculation}

%

The masses of $c\bar{c}uds$ charm pentaquark are given by
\begin{eqnarray}
 M = 2m_{u}+m_{s}+2m_{c}+ \langle P_{cs} | H | P_{cs} \rangle + C,
\end{eqnarray}
with $P_{cs}=P_{cs\mathbf{8},\mathbf{1}}$, $P'_{cs\mathbf{8},\mathbf{1}}$ and $P^{\ast}_{cs\mathbf{8},\mathbf{1}}$ and the constant term $C=C_{\Lambda} + C_{\eta_{c}}$.
The values of $a$ and $b$ in $\phi_{\mathbf{8},\mathbf{1}}$, $\phi'_{\mathbf{8},\mathbf{1}}$ and $\phi^{\ast}_{\mathbf{8},\mathbf{1}}$ are determined by the variational calculation for minimizing $\langle P_{cs} | H | P_{cs} \rangle$.

To perform the variational calculation, we need to know several matrix elements of $\vector{\lambda}_{i} \cdot \vector{\lambda}_{j}$ and $\vector{\lambda}_{i} \cdot \vector{\lambda}_{j} \, \vector{\sigma}_{i} \cdot \vector{\sigma}_{j}$ for a pair of quark $i$ and $j$ in the Hamiltonian (\ref{eq:hamiltonian_A}).
We will show the procedure of the calculations in the followings.
The color octet channel is especially important because it gives the lower energy state than the color singlet one.
In the following, therefore, we will show the matrix elements of the color octet channel, namely $P_{cs\mathbf{8}}$ ($s_{c\bar{c}}=0$), $P'_{cs\mathbf{8}}$ ($s_{c\bar{c}}=1$) and $P^{\ast}_{cs\mathbf{8}}$ ($s_{c\bar{c}}=1$).
Similar calculations can be performed for the color singlet channel.

As for $P_{cs\mathbf{8}}$, we evaluate
\begin{eqnarray}
 \langle P_{cs\mathbf{8}} | \vector{\lambda}_{q} \cdot \vector{\lambda}_{q'} | P_{cs\mathbf{8}} \rangle &=& \frac{1}{2} \Bigl( \frac{4}{3} - \frac{8}{3} \Bigr) = -\frac{2}{3}, \\
 \langle P_{cs\mathbf{8}} | \vector{\lambda}_{c} \cdot \vector{\lambda}_{\bar{c}} | P_{cs\mathbf{8}} \rangle &=& \frac{2}{3}, \\
  \langle P_{cs\mathbf{8}} | \vector{\lambda}_{c} \cdot \vector{\lambda}_{q} | P_{cs\mathbf{8}} \rangle &=& -2, \\
  \langle P_{cs\mathbf{8}} | \vector{\lambda}_{\bar{c}} \cdot \vector{\lambda}_{q} | P_{cs\mathbf{8}} \rangle &=& -2.
\end{eqnarray}
The first equation is obtained by noting that $uds$ color $\mathbf{8}$ state has the color $\bar{\mathbf{3}}$ and $\mathbf{6}$ with the same wieght.
The second equation is given by the color octet representation of $c\bar{c}$.
The last two equations are obtained transformation of the quarks, from $[c\bar{c}][qq'q'']$ to $\bar{c}q$ and $cq'q''$, as
\begin{eqnarray}
\Bigl[[c\bar{c}]_{c:\hati}[q(q'q'')_{c:\roku}]_{c:\hati}\Bigr]_{c:\iti}&=&-
\Bigl[[cq]_{c:\sanbar}[\bar{c}(q'q'')_{c:\roku}]_{c:\san}
\Bigr]_{c:\iti}\nonumber\\
&=&-\Bigl[[\bar{c}q]_{c:\hati}[c(q'q'')_{c:\roku}]_{c:\hati}
\Bigr]_{c:\iti}, \\
\Bigl[[c\bar{c}]_{c:\hati}[q(q'q'')_{c:\sanbar}]_{c:\hati}\Bigr]_{c:\iti}&=&-
\frac{1}{\sqrt{3}}\Bigl[[cq]_{c:\roku}[\bar{c}(q'q'')_{\sanbar}]_{c:\roku}
\Bigr]_{c:\iti}-\sqrt{\frac{2}{3}}\Bigl[[cq]_{\sanbar}[\bar{c}
(q'q'')_{\sanbar}]_{3}
\Bigr]_{c:\iti}\nonumber\\
&=&\frac{2\sqrt{2}}{3}\Bigl[[\bar{c}q]_{c:\iti}[c(q'q'')_{\sanbar}]_{c:\iti}
\Bigr]_{c:\iti}-\frac{1}{3}\Bigl[[\bar{c}q]_{c:\hati}[c(q'q'')_{\sanbar}]_{c:
\hati}
\Bigr]_{c:\iti},
\end{eqnarray}
with subscripts ``$c:$" the color representations.

The color-spin operators can be calculated by
\begin{eqnarray}
\braket{\po|(\lambda_q\cdot\lambda_{q'})(\sigma_q\cdot\sigma_q')|\po}
&=&\frac{1}
{2}\left[\frac{4}{3}\times1+\left(-\frac{8}{3}\right)\times(-3)\right]
=\frac{14}{3}, \\
\braket{\po|(\lambda_{c}\cdot\lambda_{\bar{c}})(\sigma_c\cdot
\sigma_{\bar{c}})|
\po}&=&\frac{2}{3}\times(-3)=-2, \\
\braket{\po|(\lambda_{c}\cdot\lambda_{q})(\sigma_c\cdot\sigma_q)+
(\lambda_{\bar{c}}\cdot\lambda_{q})(\sigma_{\bar{c}}\cdot\sigma_q)|\po} &=& 0.
\end{eqnarray}
The first equation can be obtained by noting that the symmetric states and the antisymmetric states both in spin and in color exist with the same weight in $uds$.
The second equation is trivial because $c\bar{c}$ is spin triplet.
The last equation can be obtained by changing $[\bar{c}q][cq'q'']$ or $[c\bar{c}][qq'q'']$ to $[cq][\bar{c}q'q'']$.

As for $P'_{cs\mathbf{8}}$ and $P^{\ast}_{cs\mathbf{8}}$, we perform the similar calculations for the matrix elements of $\vector{\lambda}_{i} \cdot \vector{\lambda}_{j}$ and $\vector{\lambda}_{i} \cdot \vector{\lambda}_{j} \vector{\sigma}_{i} \cdot \vector{\sigma}_{j}$.
The matrix elements of $\vector{\lambda}_{i} \cdot \vector{\lambda}_{j}$ in $P'_{cs\mathbf{8}}$ and $P^{\ast}_{cs\mathbf{8}}$ should be the same as those in $P_{cs\mathbf{8}}$.
We show the matrix elements of $\vector{\lambda}_{i} \cdot \vector{\lambda}_{j} \vector{\sigma}_{i} \cdot \vector{\sigma}_{j}$ for heavy-light $i,j$ pairs as
\begin{eqnarray}
  \langle P'_{cs\mathbf{8}} | \vector{\lambda}_{c} \cdot \vector{\lambda}_{q} \vector{\sigma}_{c} \cdot \vector{\sigma}_{q} | P'_{cs\mathbf{8}} \rangle = \frac{4}{9}, \\
  \langle P'_{cs\mathbf{8}} | \vector{\lambda}_{\bar{c}} \cdot \vector{\lambda}_{q} \vector{\sigma}_{\bar{c}} \cdot \vector{\sigma}_{q} | P'_{cs\mathbf{8}} \rangle = \frac{44}{9},
\end{eqnarray}
for $cq$ and $\bar{c}q$ pairs in $P'_{cs\mathbf{8}}$ and
\begin{eqnarray}
  \langle P^{\ast}_{cs\mathbf{8}} | \vector{\lambda}_{c} \cdot \vector{\lambda}_{q} \vector{\sigma}_{c} \cdot \vector{\sigma}_{q} | P^{\ast}_{cs\mathbf{8}} \rangle = -\frac{2}{9}, \\
  \langle P^{\ast}_{cs\mathbf{8}} | \vector{\lambda}_{\bar{c}} \cdot \vector{\lambda}_{q} \vector{\sigma}_{\bar{c}} \cdot \vector{\sigma}_{q} | P^{\ast}_{cs\mathbf{8}} \rangle = -\frac{22}{9},
\end{eqnarray}
for $cq$ and $\bar{c}q$ pairs in $P^{\ast}_{cs\mathbf{8}}$.

So far we have treated that the spin of charm quark pairs, $s_{c\bar{c}}=0$ and $s_{c\bar{c}}=1$, are conserved quantities, and regarded that $P_{cs\mathbf{8}}$ and $P'_{cs\mathbf{8}}$ (or $P_{cs\mathbf{1}}$ and $P'_{cs\mathbf{1}}$) are independent states with each other.
However, this is not necessarily correct.
It is important to comment that $P_{cs\mathbf{8}}$ and $P'_{cs\mathbf{8}}$ (or $P_{cs\mathbf{1}}$ and $P'_{cs\mathbf{1}}$) can be mixed by the color-spin mixing term $\vector{\lambda}_{i} \cdot \vector{\lambda}_{j} \vector{\sigma}_{i} \cdot \vector{\sigma}_{j}$ for a heavy (anti)quark $i$ and a light quark $j$, because both states have the common quantum number $J^{P}=1/2^{-}$ irrespective to the difference of the spin of charm quark pairs, $s_{c\bar{c}}=0$ and $s_{c\bar{c}}=1$, respectively.
The mixing effect is not so large because the spin-flip process should be suppressed by the factor $1/m_{Q}$ with the heavy quark mass $m_{Q}$,
and it can be treated as the corrections.
Therefore, we will ignore the mixing effect for simple presentation in most cases in the text, and we will treat $P_{cs\mathbf{8}}$ and $P'_{cs\mathbf{8}}$ (or $P_{cs\mathbf{1}}$ and $P'_{cs\mathbf{1}}$) as the independent states.
In the discussion part, we will consider the mixing effect for the octet case only, because it will turn out that the octet gives the ground state of the charm pentaquark $c\bar{c}uds$. 
For that purpose, we will use the matrix elements as
\begin{eqnarray}
 \braket{\pop|(\vector{\lambda}_c\cdot\vector{\lambda}_q)(\vector{\sigma}_c\cdot\vector{\sigma}_q)|\po}&=&
\frac{\sqrt{3}}{9}, \\
\braket{\pop|(\vector{\lambda}_{\bar{c}}\cdot\vector{\lambda}_q)(\vector{\sigma}_{\bar{c}}\cdot\vector{\sigma}_q)|
\po}&=&-\frac{22\sqrt{3}}{9}, \\
\braket{\pop|(\vector{\lambda}_c\cdot\vector{\lambda}_q)(\vector{\sigma}_c\cdot\vector{\sigma}_q)+(\vector{\lambda}_{\bar{c}}\cdot\vector{\lambda}_q)(\vector{\sigma}_{\bar{c}}\cdot\vector{\sigma}_q)|\po}&=&-\frac{7\sqrt{3}}{9}.
\end{eqnarray}

For the model B, it is also necessary to calculate the matrix elements of $\vector{\lambda}_{i} \cdot \vector{\lambda}_{j}$ and $\vector{\lambda}_{i} \cdot \vector{\lambda}_{j} \vector{\sigma}_{i} \cdot \vector{\sigma}_{j}$.
They are the same as those calculated for the model A.
A special attention should be paid for the three-body force in the instanton-induced interaction: it vanishes for color singlet configuration (i.e. light flavor octet) and does not vanish for the color octet configuration (i.e. light flavor singlet).

\subsection{Energy spectrum}

\begin{widetext}
\begin{figure}[tbp]
  \centering
  \includegraphics[keepaspectratio,scale=0.45]{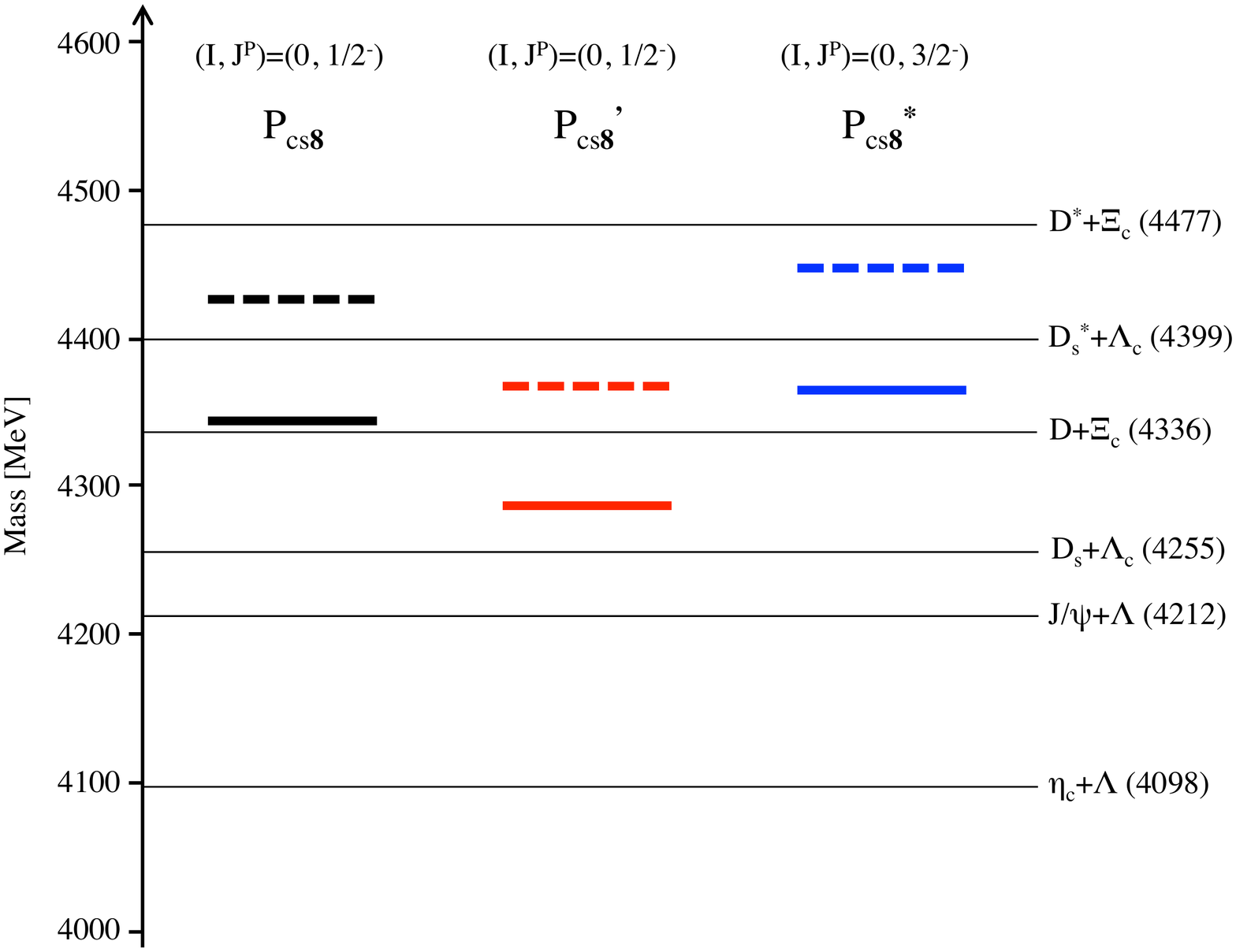}
\caption{Mass spectrum of pentaquark $P_{cs}$ with several $(I,J^{P})$ and internal color combinations ($\mathbf{8}$, $\mathbf{1}$). They are denoted by $P_{cs\mathbf{8}}$ with $c\bar{c}$ spin 0 for $(0,1/2^{-})$, $P'_{cs\mathbf{8}}$ with $c\bar{c}$ spin 1 for $(0,1/2^{-})$, and $P^{\ast}_{cs\mathbf{8}}$ with $c\bar{c}$ spin 1 for $(0,3/2^{-})$. They are colored by black, red and blue lines. The dashed lines are for the case without instanton (model A), and the solid lines are for the case with instanton (model B). The long horizontal lines are thresholds for two scattering hadrons, and the threshold energies are given in the parentheses.}
  \label{fig:mass_spectrum}
\end{figure}
\end{widetext}

The obtained numbers of the variational parameters ($a$ and $b$) and the masses of charm pentaquarks are shown in Table~\ref{table:pentaquark}.
The masses are shown also in Fig.~\ref{fig:mass_spectrum}.
Notice that the mixing between $P_{cs\mathbf{8}}$ and $P'_{cs\mathbf{8}}$ ($P_{cs\mathbf{1}}$ and $P'_{cs\mathbf{1}}$) are not considered in those results.

First, let us compare the three states $P_{cs\mathbf{1}}$ ($s_{c\bar{c}}=0$), $P'_{cs\mathbf{1}}$ ($s_{c\bar{c}}=1$) and $P^{\ast}_{cs\mathbf{1}}$ ($s_{c\bar{c}}=1$).
We notice immediately that they are much above the threshold states $\eta_{c}\Lambda$ or $J/\psi \Lambda$, and the splitting between $P_{cs\mathbf{1}}$ and $P'_{cs\mathbf{1}} \simeq P^{\ast}_{cs\mathbf{1}}$ is almost identical to the $\eta_{c}$-$J\psi$ mass difference.
This can be understood easily because in the present quark model there is no interaction between the color singlet $c\bar{c}$ and $uds$, and thus these $P_{cs\mathbf{1}}$ states are nothing but non-interacting $\eta_{c}\Lambda$ or $J/\psi \Lambda$ plus kinetic energy.
However, this simple explanation cannot applied to $P_{cs\mathbf{8}}$, $P'_{cs\mathbf{8}}$ and $P^{\ast}_{cs\mathbf{8}}$ due to the complicated color structure.

Second, one of the most interesting observations is that, in color octet, the instanton-induced interaction reduces very much the mass of charm pentaquarks than the one-gluon exchange, while there is no large change in color singlet.
Let us understand why the large reduction of mass in color octet arises.
Based on the above observation, one may expect that the mass reduction in color octet is in fact supplied by the instanton-induced interaction.
However, the actual mechanism may not be so simple.
We can check the attraction and repulsion of the instanton-induced interaction by decomposing the matrix elements of the Hamiltonian.
Then, we find that the two-body interaction part ($V_{\mathrm{III}2}$ in Eq.~(\ref{eq:III2_QM})) gives an attraction, while the three-body part ($V_{\mathrm{III}3}$ in Eq.~(\ref{eq:III3_QM})) gives a repulsion.
Because the $uds$ flavor is singlet in color octet channel, the anti-symmetry of any two pairs of quarks gives a strong attraction in $V_{\mathrm{III}2}$.
In fact, the attraction in color octet ($uds$ singlet) is stronger than the attraction in color singlet ($uds$ octet).
At the same time, however, it give also a strong repulsion in $V_{\mathrm{III}3}$.
As a result, the attraction in $V_{\mathrm{III}2}$ is almost canceled by the repulsion in $V_{\mathrm{III}3}$, and hence the instanton-induced interaction does not provide much attraction.
We should consider rather that the attraction is mainly provided by the one-gluon exchange rather than the instanton-induced interaction.

\begin{table}[tbp]
	\begin{center}
		\caption{Masses of $P_{c}^{\mathbf{1}_{\!f}}$ ($M$) with several $(I,J^{P})$ and color combinations: $P_{cs\mathbf{8}}$ and $P_{cs\mathbf{1}}$ with $c\bar{c}$ spin 0 for $(0,1/2^{-})$, $P'_{cs\mathbf{8}}$ and $P'_{cs\mathbf{1}}$ with $c\bar{c}$ spin 1 for $(0,1/2^{-})$, and $P^{\ast}_{cs\mathbf{8}}$ and $P^{\ast}_{cs\mathbf{1}}$ with $c\bar{c}$ spin 1 for $(0,3/2^{-})$. The determined values of $a$ and $b$ are displayed also. The model A contains the one-gluon exchange only at short distance force, and the model B contains both the one-gluon exchange and the instanton-induced interaction.}
		\begin{tabular}{|c|c|c|c|c|c|c|c|}\hline
			 \multicolumn{2}{|c}{($I,J^{P}$)} & \multicolumn{2}{|c|}{$(1,1/2^{-})$} &  \multicolumn{2}{|c|}{$(1,1/2^{-})$} & \multicolumn{2}{|c|}{$(1,3/2^{-})$} \\
			 \cline{1-8}
                          \multicolumn{2}{|c|}{color configuration} & $P_{cs\mathbf{8}}$ & $P_{cs\mathbf{1}}$ & $P'_{cs\mathbf{8}}$ & $P'_{cs\mathbf{1}}$ & $P^{\ast}_{cs\mathbf{8}}$ & $P^{\ast}_{cs\mathbf{1}}$ \\ \hline \hline
                   & $M$ [MeV] & 4427.2 & 4400.2 & 4366.6 & 4512.2 & 4448.2 & 4512.2 \\ \cline{2-8}
     model A & $a$ [fm] & 0.331 & 0.198 & 0.313 & 0.258 & 0.334 & 0.258 \\ \cline{2-8}
                   & $b$ [fm] & 0.511 & 0.542 & 0.492 & 0.542 & 0.518 & 0.542 \\ \hline \hline
                   & $M$ [MeV] & 4343.8 & 4409.3 & 4286.4 & 4512.3 & 4363.7 & 4512.3 \\ \cline{2-8}
     model B & $a$ [fm] & 0.333 & 0.198 & 0.316 & 0.258 & 0.336 & 0.258 \\ \cline{2-8}
                   & $b$ [fm] & 0.521 & 0.540 & 0.505 & 0.540 & 0.528 & 0.540 \\ \hline
		\end{tabular}
		\label{table:pentaquark}
	\end{center}
\end{table}

\begin{widetext}
\begin{figure}[tbp]
  \centering
  \includegraphics[keepaspectratio,scale=0.3]{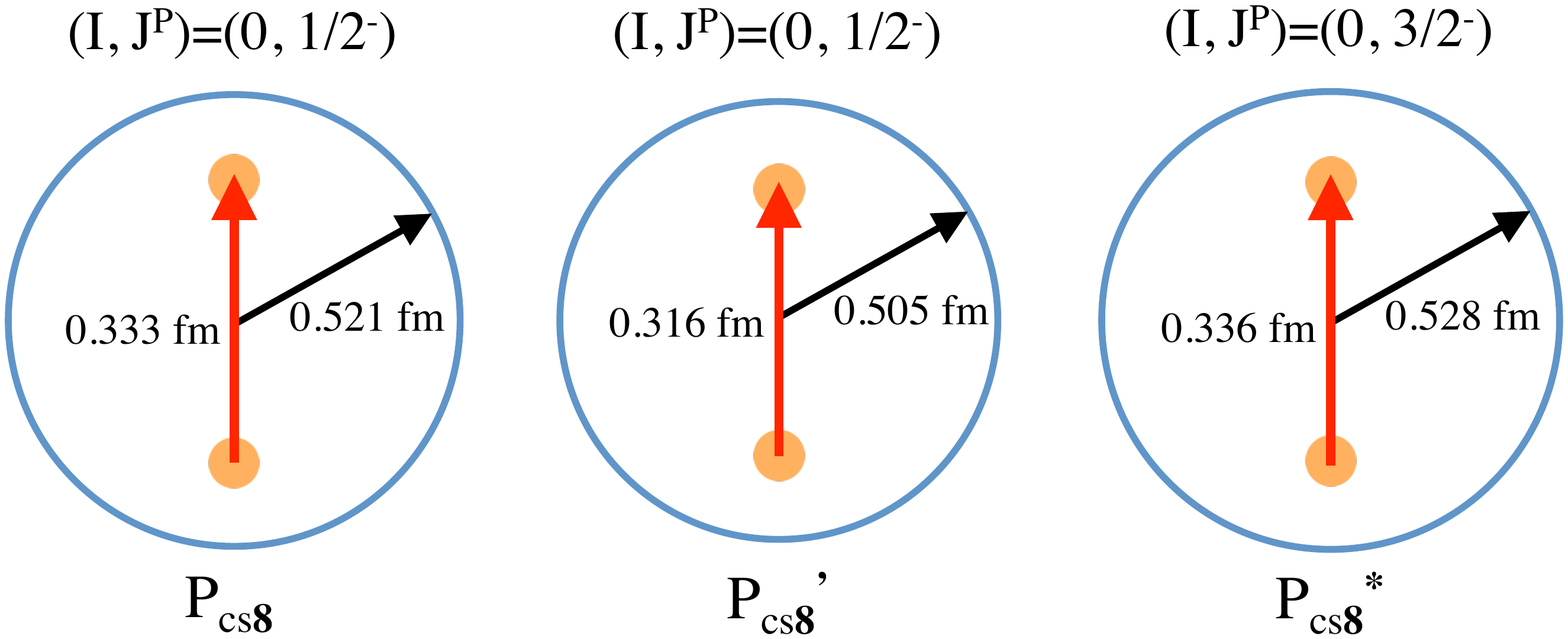}
  \vspace{-4em}
\caption{The diagrams for the obtained values of the variational parameters $a$ (red thick arrow) and $b$ (thin black arrow) in the model B for $P_{cs\mathbf{8}}$, $P'_{cs\mathbf{8}}$ and $P^{\ast}_{cs\mathbf{8}}$ (cf.~Table~\ref{table:pentaquark}).}
  \label{fig:size}
\end{figure}
\end{widetext}

It is interesting to compare the size of inter-quark distance for color-octet configuration $P_{cs\mathbf{8}}$, $P'_{cs\mathbf{8}}$, $P^{\ast}_{cs\mathbf{8}}$) and color-singlet configuration ($P_{cs\mathbf{1}}$, $P'_{cs\mathbf{1}}$, $P^{\ast}_{cs\mathbf{1}}$).
As a general tendency, in Table.~\ref{table:pentaquark}, we notice that the sizes between $c$ and $\bar{c}$ ($a$) in color-octet configuration is larger than those in color-singlet configuration.
This behavior can be understood in the following way.
The important role is played by the $\vector{\lambda}_{i} \cdot \vector{\lambda}_{j}$ operators, which are included in the color Coulomb potential and the linear confinement potential.
As for the $c\bar{c}$ potential, we find $\vector{\lambda}_{c} \cdot \vector{\lambda}_{\bar{c}}=2/3$ for color-octet configuration ($P_{cs\mathbf{8}}$, $P'_{cs\mathbf{8}}$, $P^{\ast}_{cs\mathbf{8}}$) and $\vector{\lambda}_{c} \cdot \vector{\lambda}_{\bar{c}}=-16/3$ for color-singlet configuration ($P_{cs\mathbf{1}}$, $P'_{cs\mathbf{1}}$, $P^{\ast}_{cs\mathbf{1}}$).
Due to the repulsion and attraction in each configuration, the $c\bar{c}$ sizes in color-octet are larger than those in color-singlet (see Fig.~\ref{fig:size}).
On the other hand, the sizes of wave functions of light quarks ($b$) in color-octet configuration is smaller than those in color-singlet configuration.
This is also understood from the values of $\vector{\lambda}_{i} \cdot \vector{\lambda}_{j}$, though the situation is a bit cumbersome.
When we compare the value of $\vector{\lambda}_{q} \cdot \vector{\lambda}_{q}$ for a pair of light quarks, we find from Table~\ref{table:color} that both color-octet and -singlet configurations feel attraction provided that the former attraction is less attractive.
Hence we may expect that the size of $b$ in color-octet is larger than that in color-singlet.
However, this is not the case.
The trick is that the attraction by $c$ ($\bar{c}$) and $q$, $\vector{\lambda}_{c} \cdot \vector{\lambda}_{q}$ ($\vector{\lambda}_{\bar{c}} \cdot \vector{\lambda}_{q}$), exists only for color-octet configuration.
This provides the shrinkage of the wave function of the light quarks in color-octet configuration (Fig.~\ref{fig:size}).

\begin{table}[hbtp]
	\begin{center}
				\caption{Expectation values for octet-type configuration ($P_{cs\mathbf{8}}$, $P'_{cs\mathbf{8}}$, $P^{\ast}_{cs\mathbf{8}}$) and singlet-type configuration ($P_{cs\mathbf{1}}$, $P'_{cs\mathbf{1}}$, $P^{\ast}_{cs\mathbf{1}}$).}
		\begin{tabular}{|c|c|c|}\hline
			& octet type & singlet type \\ \hline
			\hline
			$\vector{\lambda}_c \cdot \vector{\lambda}_{\bar{c}}$ & $\frac{2}{3}$ & $-\frac{16}{3}$ \\ \hline
			$\vector{\lambda}_q \cdot \vector{\lambda}_{q}$ & $-\frac{2}{3}$ & $-\frac{8}{3}$ \\ \hline
			$\vector{\lambda}_c \cdot \vector{\lambda}_{q}$ & -2 & 0 \\ \hline
			$\vector{\lambda}_{\bar{c}} \cdot \vector{\lambda}_{q}$ & -2 & 0 \\ \hline
		\end{tabular}
		\label{table:color}
	\end{center}
\end{table}

\begin{figure}[tbp]
  \centering
  \includegraphics[keepaspectratio,scale=0.25]{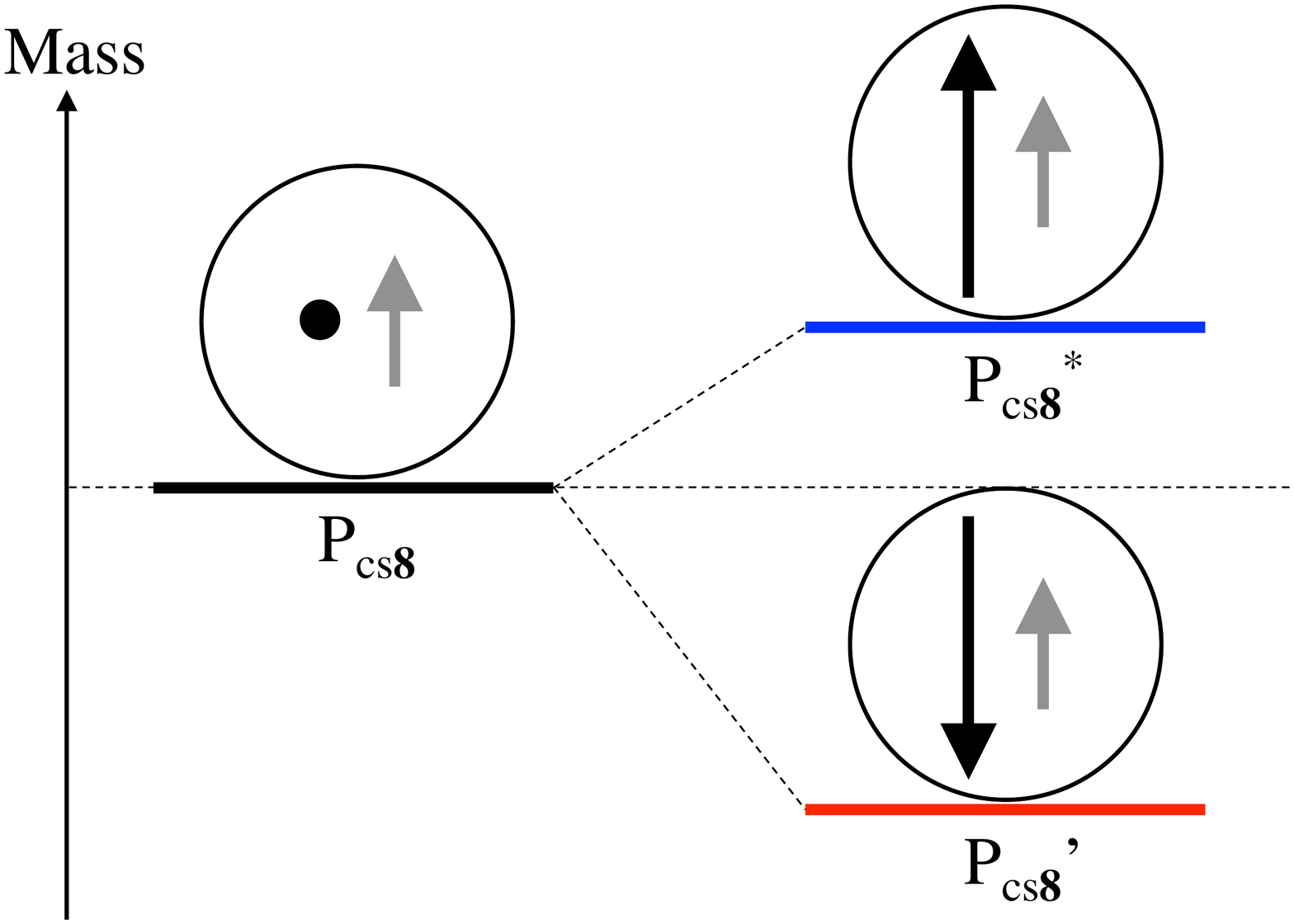}
\caption{The splitting of mass spectrum of $P_{cs\mathbf{8}}$, $P'_{cs\mathbf{8}}$ and $P^{\ast}_{cs\mathbf{8}}$. The left black blob and arrows in the circle indicate the $c\bar{c}$ spin 0 and 1, respectively, and the right gray arrows do the spin 1/2 of $uds$ component. }
  \label{fig:mass_split}
\end{figure}

In Fig.~\ref{fig:mass_spectrum}, we notice that the masses of $P_{cs\mathbf{8}}$, $P'_{cs\mathbf{8}}$ and $P^{\ast}_{cs\mathbf{8}}$, $M_{P_{cs\mathbf{8}}}$, $M_{P'_{cs\mathbf{8}}}$ and $M_{P^{\ast}_{cs\mathbf{8}}}$, are in order as given by
\begin{eqnarray}
 M_{P'_{cs\mathbf{8}}} < M_{P_{cs\mathbf{8}}} < M_{P^{\ast}_{cs\mathbf{8}}},
\end{eqnarray}
both for the model A and the model B.
This is naturally understood from the color-spin interaction part containing $\vector{s}_{i} \cdot \vector{s}_{j}$ part.
We consider the color clusters $c\bar{c}$ with color octet and spin 0 or 1 and $uds$ with color octet and spin 1/2.
When the $c\bar{c}$ cluster has spin 0, there is no spin-spin interaction.
When the $c\bar{c}$ cluster has spin 1, the compound states $c\bar{c}uds$ are split to the two states with total spin 3/2 and 1/2.
The spin-spin operator gives the energy splitting for those two states, a repulsion for the former and an attraction for the latter (the strength fraction two-to-one), and hence the masses become different as shown in Fig.~\ref{fig:mass_split}

\subsection{Mixing between $P_{cs\mathbf{8}}$ and $P_{cs\mathbf{8}}'$}

Up to now, we have neglected the mixing of $P_{cs\mathbf{8}}$ and $P'_{cs\mathbf{8}}$.
The mixing interaction is suppressed by the factor $1/m_{c}^{2}$ in the spin-spin interaction in Eq.~(\ref{eq:CMI}) because the former contains the $c\bar{c}$ spin $s_{c\bar{c}}=0$ and the latter does $s_{c\bar{c}}=1$, and hence to ignore the mixing is a good approximation.
We will investigate the accuracy of this approximation by considering the mixing of $P_{cs\mathbf{8}}$ and $P'_{cs\mathbf{8}}$.
In this case, we consider the superposed state
\begin{eqnarray}
 \ket{\Psi}=c_1\ket{\po}+c_2\ket{\pop},
\end{eqnarray}
with coefficients $c_{1}$ and $c_{2}$.
The Schr\"odinger equation is schematically expressed as
\begin{eqnarray}
\left(\begin{array}{cc}
\braket{\po|H|\po}&\braket{\po|H|\pop}\\
\braket{\pop|H|\po}&\braket{\pop|H|\pop}
\end{array}\right)\left(\begin{array}{c}
c_1\\
c_2
\end{array}\right)=E\left(\begin{array}{c}
c_1\\
c_2
\end{array}\right).
\end{eqnarray}
The energy $E$ as an eigenvalue is given by $E_{L}$ for lower energy and by $E_{H}$ for higher energy,
\begin{eqnarray}
 E_L 
 &=&\frac{1}{2}\left(\braket{\po|H|\po}+\braket{\pop|H|\pop}-
\sqrt{\left(\braket{\po|H|\po}-\braket{\pop|H|\pop}\right)^2+4 \left| \braket{\pop|H|\po} \right|^2}\right),
\nonumber\\
\label{mixsita}\\
E_H 
&=&\frac{1}{2}\left(\braket{\po|H|\po}+\braket{\pop|H|\pop}+
\sqrt{\left(\braket{\po|H|\po}-\braket{\pop|H|\pop}\right)^2+4 \left| \braket{\pop|H|\po} \right|^2}\right),
\nonumber\\
\label{mixsitb}
\end{eqnarray}
and the corresponding states will be denoted by $P_{cs\mathbf{8}}^{L}$ and $P_{cs\mathbf{8}}^{H}$, respectively.
In the variational calculation to obtain $E_{L}$ and $E_{H}$, we use different size parameters in the spatial parts in the wave functions, $(a_{1}, b_{1})$ for $P_{cs\mathbf{8}}$ and $(a_{2}, b_{2})$ for $P'_{cs\mathbf{8}}$.
However, we find that $(a_{1}, b_{1})$ are only slightly different from $(a_{2}, b_{2})$; $a_{1}=0.309 \,\mathrm{fm}, b_{1}=0.483 \,\mathrm{fm}$ and $a_{2}=0.307 \,\mathrm{fm}, b_{2}=0.484 \,\mathrm{fm}$ in the model A, $a_{1}=0.313 \,\mathrm{fm}, b_{1}=0.498 \,\mathrm{fm}$ and $a_{2}=0.311 \,\mathrm{fm}, b_{2}=0.499 \,\mathrm{fm}$ in the model B.
The obtained energy $E_{L}$ and $E_{H}$ as well as the fractions of $P_{cs\mathbf{8}}$ component and $P'_{cs\mathbf{8}}$ component are shown in Table~\ref{table:mixwithout} and \ref{table:mixwith}.
Comparing the results of the masses of $P_{cs\hati}$ and $P'_{cs\hati}$ summarized in Table~\ref{table:pentaquark}, we find that the mass of $P_{cs\hati}^{L}$ becomes smaller by about 20 MeV and $P'_{cs\hati}$ becomes larger by about 30 MeV.
The mixing fractions are about 20 \%.
This value is consistent with the results in Ref.~\cite{Yuan:2012wz}.
In this reference the state corresponding to ours is supplied by the combinations of $|1'\rangle$ and $|3'\rangle$ in $[211]$ state, which contains a flavor singlet state.
Notice that $c\bar{c}$ spin $s_{c\bar{c}}=0,1$ are mixed in each of $|1'\rangle$ and $|3'\rangle$s.

\begin{table}[tbp]
	\begin{center}
		\caption{Energy $E_{L}$ and $E_{H}$ and fractions of $\po$ and $\pop$ after mixing of $\po$ and $\pop$ in the model A.}
			\begin{tabular}{|c||c|c|c|}\hline
				$J^P=1/2^-$ mixing state& Energy [$\mathrm{MeV}$] & 
				$\po$ fraction [\%] & 
				$\pop$ fraction [\%]\\ 
				\hline\hline
				$P_{cs\hati}^{L}$ (lower state) & 4343.0 & 79.2 & 20.8 \\ \hline
				$P_{cs\hati}^{H}$ (higher state) & 4459.6 & 20.8 & 79.2 \\ \hline
			\end{tabular}\label{table:mixwithout}
		\end{center}
	\end{table}

	\begin{table}[tbp]
		\begin{center}
			\caption{Energy $E_{L}$ and $E_{H}$ and fractions of $\po$ and $\pop$ after mixing of $\po$ and $\pop$ in the model B.}
			\begin{tabular}{|c||c|c|c|}\hline
				$J^P=1/2^-$ mixing state& Energy [$\mathrm{MeV}$] & 
				$\po$ fraction [\%] & 
				$\pop$ fraction [\%]\\ 
				\hline\hline
				$P_{cs\hati}^{L}$ (lower state) & 4264.5 & 79.3 & 20.7 \\ \hline
				$P_{cs\hati}^{H}$ (higher state) & 4372.5 & 20.7 & 79.3 \\ \hline
			\end{tabular}\label{table:mixwith}
		\end{center}
	\end{table}

\section{Discussion}

We investigate the possible decay modes of the charm pentaquark $P_{cs\mathbf{8}}$, $P'_{cs\mathbf{8}}$ and $P^{\ast}_{cs\mathbf{8}}$ in the model B.
The obtained masses are located above thresholds of several open channels, as shown in Fig.~\ref{fig:mass_spectrum}.
The available decay channels are $\eta_{c}+\Lambda$, $J/\psi+\Lambda$, $D_{s}+\Lambda_{c}$ and $D+\Xi_{c}$.
The most lowest threshold is given by $\eta_{c}+\Lambda$, and the next lowest is by $J/\psi+\Lambda$.
However, those two decay channels are suppressed by the effect of the light flavor SU(3)$_{\mathrm{f}}$ breaking and the heavy quark HQS breaking.
Because $P_{cs\mathbf{8}}$, $P'_{cs\mathbf{8}}$ and $P^{\ast}_{cs\mathbf{8}}$ are flavor singlet, the decay to $\eta_{c}+\Lambda$ and/or $J/\psi+\Lambda$ breals SU(3)$_{\mathrm{f}}$ symmetry.
Concerning $P_{cs\mathbf{8}}$, the decay to $J/\psi+\Lambda$ is further suppressed by the HQS breaking, because the spin of $c\bar{c}$ pair in $P_{cs\mathbf{8}}$ is predominantly singlet.
Concerning $P'_{cs\mathbf{8}}$ and $P^{\ast}_{cs\mathbf{8}}$, in contrast, the decay to $\eta_{c}+\Lambda$ is suppressed by the HQS breaking, because the spins of $c\bar{c}$ pair in $P_{cs\mathbf{8}}$ are approximately triplet.
For $P_{cs\mathbf{8}}$, $P'_{cs\mathbf{8}}$ and $P^{\ast}_{cs\mathbf{8}}$, the decays to $D_{s}+\Lambda_{c}$ and $D+\Xi_{c}$ are not suppressed both in the SU(3)$_{\mathrm{f}}$ and in the HQS breaking.
Though there may be some contributions which are not neglected for $P_{cs\mathbf{8}}$, $P'_{cs\mathbf{8}}$ because of S-wave decay,
it may be possible that the emission energy is not so large, and hence the small phase space may make the decay widths small.
The decay from $P^{\ast}_{cs\mathbf{8}}$ (spin 3/2) is expected to be suppressed because it is D-wave decay.

We may consider the three-body state in the final state.
The example is given by $\eta_{c}+\pi+\Sigma$ (threshold energy 4315 MeV).
This decay process is not suppressed by the SU(3)$_{\mathrm{f}}$ breaking.
However, the phase space of the three-body final states is smaller than that in two-body final sate, and hence the decay widths may not be so large.
We may also consider that the decay widths could be suppressed because the color degrees of freedom should be recombined from the color octet in the initial state to the final state $\eta_{c}+\Lambda$ and $J/\psi+\Lambda$.
To estimate the decay widths quantitatively is left as future works.

\begin{table}[htbp]
\caption{Possible decay modes of charm pentaquark $P_{cs\mathbf{8}}$, $P'_{cs\mathbf{8}}$ and $P^{\ast}_{cs\mathbf{8}}$ in the model B (cf.~Fig.~\ref{fig:mass_spectrum}). The decays to SU(3)$_{\mathrm{f}}$ singlet final state is suppressed as indicated by ``SU(3)$_{\mathrm{f}}$", because $P_{cs\mathbf{8}}$, $P'_{cs\mathbf{8}}$ and $P^{\ast}_{cs\mathbf{8}}$ are SU(3)$_{\mathrm{f}}$ octet. The decay to the final state including $\eta_{c}$ ($J/\psi$) is suppressed for the initial state $P'_{cs\mathbf{8}}$ and $P^{\ast}_{cs\mathbf{8}}$ with $s_{c\bar{c}}=1$ ($P_{cs\mathbf{8}}$ with $s_{c\bar{c}}=0$), as denoted by ``HQS". The decay channels in the last two rows are suppressed by the color recombination (``color recomb.").}
\begin{center}
\begin{tabular}{|c|c|c|c|}
\hline
 Decay channels  & $P_{cs\mathbf{8}}$ ($s_{c\bar{c}}\!=\!0$) & $P'_{cs\mathbf{8}}$ ($s_{c\bar{c}}\!=\!1$) & $P^{\ast}_{cs\mathbf{8}}$ ($s_{c\bar{c}}\!=\!1$) \\
\hline
 $\eta_{c}+\Lambda$ & SU(3)$_{\mathrm{f}}$ & SU(3)$_{\mathrm{f}}$ and HQS & SU(3)$_{\mathrm{f}}$ and HQS \\
\hline
 $J/\psi+\Lambda$ & SU(3)$_{\mathrm{f}}$ and HQS & SU(3)$_{\mathrm{f}}$ & SU(3)$_{\mathrm{f}}$ \\
\hline
 $D_{s}+\Lambda_{c}$ & color recomb. & color recomb. & color recomb. \\
\hline
 $D+\Xi_{c}$ & color recomb. & color recomb. & color recomb. \\
\hline
\end{tabular}
\end{center}
\label{default}
\end{table}%

\section{Conclusion}
We investigate the internal structure of $c\bar{c}uds$ charm pentaquark, in which $c\bar{c}$ cluster is the color octet state.
This is an exotic color configuration which cannot be realized in charmonia.
The light flavor multiplet of this state is flavor-singlet.
By adopting the color-spin interaction and the instanton-induced interaction, we have found that $P_{cs\bf{8}}'$ with total spin $1/2$ and $c\bar{c}$ spin 1 will be the most stable state, while the other states, $P_{cs\bf{8}}$ with total spin $1/2$ and $c\bar{c}$ spin 0 and $P_{cs\bf{8}}^{\ast}$ with total spin $3/2$ and $c\bar{c}$ spin 1, are the excited states.
The size of $c\bar{c}$ as well as the size of $uds$ in those states are much less than one fm, and hence they are the compact multiquark states.
We investigate also the mixing of the $P_{cs\bf{8}}$ and $P_{cs\bf{8}}'$ due to the breaking of the heavy quark symmetry, but find that the mixing effect is not so large.
We discuss several possible decay process of $c\bar{c}uds$ for the obtained masses, and find many channels should be suppressed by light flavor SU(3) symmetry or by the heavy quark symmetry or by both of them.
Therefore, we conclude that the $c\bar{c}uds$ pentaquark is a candidate which should be searched in experimental studies.
This is an interesting subject for experiments at high energy accelerator facilities.

\section*{Acknowledgments}
This work is supported by the Grant-in-Aid for Scientific Research (Grant No.~JP25247036, No.~JP15K17641 and No.~JP16K05366) from Japan Society for the Promotion of Science (JSPS).

\end{document}